\documentclass{article}
\usepackage{spconf,amsmath,graphicx}
\usepackage{amsfonts}
\usepackage{subcaption}
\usepackage{tikz}
\usetikzlibrary{calc,chains,shapes,positioning}
\usepackage{comment}
\usepackage{pgfplots}
\usepackage{phaistos}
\usepackage{float}
\usepackage{phaistos}
\usepackage{pgfplots}
\usepackage{cite}
\usepackage{mathdots}
\usepackage{scalerel}
\usepackage{enumitem}
\usepackage[ruled]{algorithm2e}
\SetAlgoHangIndent{0pt}
\usepackage[subtle,title=normal,sections=normal,margins=normal,bibliography=normal,mathdisplays=normal,mathspacing=normal]{savetrees}
\usepackage{balance}
\usepackage{epstopdf}

\SetCommentSty{mycommfont}

\title{Microphone Subset Selection for the Weighted Prediction Error Algorithm\\ using a Group Sparsity Penalty}
\name{Anselm Lohmann$^{1}$, Toon van Waterschoot$^{2}$, Joerg Bitzer$^{3}$, Simon Doclo$^{1,3}$ \thanks{This work has received funding from the European Union’s Horizon 2020 research and innovation programme under the Marie Skłodowska-Curie grant
agreement No. 956369 and the ERC Consolidator Grant: SONORA (no. 773268), and from the Deutsche Forschungsgemeinschaft (DFG, German Research Foundation) under Germany's Excellence Strategy - EXC 2177/1 - Project ID 390895286.}}
\address{$^{1}$University of Oldenburg, Dept. of Medical Physics and Acoustics and Cluster of Excellence Hearing4all, Germany \\
$^{2}$KU Leuven, Department of Electrical Engineering (ESAT-STADIUS), Leuven, Belgium\\
$^{3}$Fraunhofer IDMT, Project Group Hearing, Speech and Audio Technology, Oldenburg, Germany \\
{\tt anselm.lohmann@uni-oldenburg.de}
}

\begin{document}
\ninept
\maketitle
\begin{abstract}
Reverberation can severely degrade the quality of speech signals recorded using microphones in an enclosure. In acoustic sensor networks with spatially distributed microphones, a similar dereverberation performance may be achieved using only a subset of all available microphones. Using the popular convex relaxation method, in this paper we propose to perform microphone subset selection for the weighted prediction error (WPE) multi-channel dereverberation algorithm by introducing a group sparsity penalty on the prediction filter coefficients. The resulting problem is shown to be solved efficiently using the accelerated proximal gradient algorithm. Experimental evaluation using measured impulse responses shows that the performance of the proposed method is close to the optimal performance obtained by exhaustive search, both for frequency-dependent as well as frequency-independent microphone subset selection. Furthermore, the performance using only a few microphones for frequency-independent microphone subset selection is only marginally worse than using all available microphones.
\end{abstract}
\section{Introduction}
\label{sec:intro}
Microphone recordings of a speech source inside an enclosure are typically degraded by reverberation, i.e. acoustic reflections against walls and objects in the enclosure. While early reflections may improve speech intelligibility, late reverberation typically reduces both speech intelligibility as well as automatic speech recognition performance \cite{Beutelmann2006,YoshiokaASR2012}. Therefore, effective speech dereverberation is required for many applications, including voice-controlled systems, hearing aids and hands-free telephony \cite{HabetsNaylor2018,Cauchi2015,Braun2018,Dietzen2019,Li2019,Williamson2017,Lemercier2022,Nakatani2010,Jukic2015,Wung2020,Witkowski2021,Huang2022}. A popular blind multi-channel dereverberation algorithm is the weighted prediction error (WPE) algorithm \cite{Nakatani2010,Jukic2015,Wung2020,Witkowski2021,Huang2022}, which is based on multi-channel linear prediction (MCLP). WPE performs dereverberation by estimating a multi-channel prediction filter to predict the late reverberant component in a reference microphone and subtracting this estimate from the reference microphone signal. Several variants of the WPE algorithm have been proposed, e.g., aiming at controlling sparsity of the dereverberated output signal in the time-frequency domain \cite{Jukic2015,Witkowski2021}. \par
In multi-microphone processing for compact arrays, typically all available microphones are utilized. However, when considering spatially distributed microphones, the spatial diversity of the microphone signals may allow for similar performance using only a subset of microphones, reducing computational complexity. However, microphone subset selection is a combinatorial problem, which may become computationally infeasible when using a large number of microphones. Several microphone subset selection methods have been proposed for different speech enhancement algorithms, e.g., beamforming \cite{Szurley2012, Zhang2018,Afifi2021,Hu2023}. However, to the best of our knowledge no microphone subset selection method for the WPE algorithm exists. 

Using the popular convex relaxation approach \cite{Joshi2009}, in this paper we propose to perform microphone subset selection for the WPE algorithm by introducing a group sparsity penalty on the prediction filter coefficients. The group sparsity penalty helps promote a sparse representation among the filter coefficients for different groups, i.e. microphones, which has proven effective for subset selection \cite{Li2017}. The resulting problem is shown to be solved efficiently using the accelerated proximal gradient algorithm. In the proposed method, first a group-sparse prediction filter is computed using the fast iterative shrinkage thresholding algorithm (FISTA) to then select the microphones with the largest prediction filter coefficients in the $\ell_2$-norm sense in a variable selection step. The proposed method is evaluated using measured impulse responses for 9 spatially distributed microphones in a measurement laboratory \cite{Fejgin2023} with a reverberation time $T_{60}$ of approximately 1300 ms for different source positions. The results show that the performance of the proposed method is close to the optimal performance using exhaustive search for both for frequency-dependent as well as frequency-independent microphone subset selection for a suitable choice of the group sparsity factor. Furthermore, even when performing frequency-independent microphone subset selection with a fixed group sparsity factor, the performance using the resulting subset of microphones is only marginally worse than using all microphones.

\section{Signal Model}
\label{sec:format}
We consider a scenario where a single speech source is captured in an enclosure by $M$ spatially-distributed microphones. Similarly as in \cite{Nakatani2010,Jukic2015,Witkowski2021}, we consider a scenario without additive noise. In the short time Fourier transform (STFT)-domain, let $s (f,n)$ denote the clean speech signal with $f \in \{1,...,F\}$ the frequency bin index and $n \in \{1,...,N\}$ the time frame index, where $F$ and $N$ denote the number of frequency bins and time frames respectively. The reverberant signal at the $m$-th microphone $x_m (f,n)$ can be written as
\begin{equation} 
    x_m (f,n) = \sum_{l = 0}^{L_h - 1}h_m(f,l)s(f,n-l) + e_m (f,n),
    \label{eq:sigmodel_stft}
\end{equation}
where $h_m (f,n)$ denotes the subband convolutive transfer function with length $L_h$ between the speech source and the $m$-th microphone, and $e_m(f,n)$ denotes the subband modelling error \cite{Avargel2007}.  Without loss of generality, we define the first microphone as the reference microphone. Assuming the term $e_m (f,n)$ in \eqref{eq:sigmodel_stft} can be disregarded, the dereverberation problem, with the index $f$ omitted, can be formulated as
\begin{equation}
    d(n) = x_1(n) - r(n).
    \label{eq:sig_model}
\end{equation}
The desired component $d(n) = \sum_{l = 0}^{L_d - 1}h_1(l)s(n-l)$ consists of the direct path and early reflections in the reference microphone signal $x_1(n)$, where $L_d$ denotes the temporal cut-off between early and late reflections. The undesired component $r(n) = \sum_{l = L_d}^{L_h - 1}h_1(l)s(n-l)$, which we aim to estimate, is the late reverberant component in the reference microphone signal $x_1(n)$. Using the MCLP model \cite{Nakatani2010}, the late reverberant component $r(n)$ can be written as the sum of delayed filtered versions of all reverberant microphone signals. Whereas for compact microphone arrays typically the same prediction delay is used in each microphone, it has recently been shown in \cite{Lohmann2023} that for spatially distributed microphones it is beneficial to use a microphone-dependent prediction delay, i.e.
\begin{equation}
r(n) = \sum_{m = 1}^{M}\sum_{l=0}^{L_g - 1}g_m(l)x_m(n - \tau_m - l),
\label{eq:new_MCLP}
\end{equation}
where $g_m(l)$ denotes the $m$-th prediction filter of length $L_g$ and $\tau_m$ denotes the prediction delay for the $m$-th microphone.
Using \eqref{eq:new_MCLP}, the signal model in \eqref{eq:sig_model} can be rewritten in vector notation  as
\begin{equation}
    \mathbf{d} = \mathbf{x}_1 - \mathbf{X}_{\boldsymbol{\tau}}\mathbf{g},
    \label{eq:new_vec_MCLP}
\end{equation}
with
\begin{equation}
\mathbf{d} = \begin{bmatrix} 
d(1)&\cdots & d (N)
\end{bmatrix}^{T} \in\mathbb{C}^{N},
\label{eq:d_def}
\end{equation}
\begin{equation}
\mathbf{x}_1 = \begin{bmatrix} 
x_1(1)&\cdots & x_1(N) 
\end{bmatrix}^{T} \in\mathbb{C}^{N}.
\end{equation}
The multi-channel delayed convolution matrix $\mathbf{X}_{\boldsymbol{\tau}}$ in \eqref{eq:new_vec_MCLP} is defined as
\begin{equation}
\mathbf{X}_{\boldsymbol{\tau}} = \begin{bmatrix} 
\mathbf{X}_{\tau_1}&\cdots & \mathbf{X}_{\tau_{M}}
\end{bmatrix}\in\mathbb{C}^{N\times ML_g},
\end{equation} 
where $\mathbf{X}_{\tau_m} \in\mathbb{C}^{N\times L_g}$ is the convolution matrix of $\mathbf{x}_m$ delayed by $\tau_m$ frames with $\tau = \tau_1$ the prediction delay in the reference microphone. The prediction filter $\mathbf{g}$ is defined as
\begin{equation}
\mathbf{g} = \begin{bmatrix} 
\mathbf{g}_{1}^T&\cdots & \mathbf{g}_{M}^T
\end{bmatrix}^T\in\mathbb{C}^{ML_g},
\label{eq:filter_g}
\end{equation} 
where $\mathbf{g}_m \in\mathbb{C}^{L_g}$ is the stacked vector of the filter coefficients $g_m(n)$.
\section{Microphone Subset Selection}
In this section, we propose a method to perform microphone subset selection for the WPE algorithm. Using the convex relaxation approach, we perform microphone subset selection by introducing a group sparsity penalty on the prediction filter coefficients. The resulting problem is shown to be efficiently solved using the accelerated proximal gradient algorithm. After computing the group-sparse prediction filter, we perform a variable selection step, which is typical for convex relaxation based methods. In Section 3.1, we first define the combinatorial microphone subset selection problem using the $\ell_0$-norm and perform convex relaxation to reformulate the nonconvex combinatorial problem. In Section 3.2, we discuss the solution of the resulting problem using the proximal gradient algorithm. In Section 3.3, we discuss the variable selection step on the computed group-sparse prediction filter.
\subsection{Convex relaxation}
In \cite{Jukic2015}, it has been shown that the WPE problem can be reformulated as an $\ell_p$-norm minimization problem
\begin{equation}
\min_{\mathbf{g}}J(\mathbf{g}) = \left\| \mathbf{d}\right\|_p = \left\| \mathbf{x_1} - \mathbf{X}_{\boldsymbol{\tau}}\mathbf{g} \right\|_p,
\label{ref:og_problem}
\end{equation}
where $\left\|.\right\|_p$ denotes the $\ell_p$-norm. For effective dereverberation, the sparsity-promoting parameter $p$ is typically chosen in the range $0<p<1$ \cite{Jukic2015}, leading to a nonconvex optimization problem in \eqref{ref:og_problem}. When selecting a (frequency-dependent) subset $\mathcal{S}$ of $K < M$ microphones, $M-K$ groups $\mathbf{g}_m$ of the prediction filter $\mathbf{g}$ in \eqref{eq:filter_g} need to be set to the zero vector. Since the reference microphone always needs to be part of the subset $\mathcal{S}$, this can be reformulated as 
\begin{equation}
\left\|\mathbf{u}\right\|_0 = K-1,
\label{ref:l0_equality}
\end{equation}
where $\left\|.\right\|_0$ denotes the $\ell_0$-norm and $\mathbf{u}$ denotes the group vector, which contains the $\ell_2$-norms of the prediction filter groups $\mathbf{g}_m$ (not including the reference microphone), i.e.
\begin{equation}
\mathbf{u} = 
\begin{bmatrix} \left\|\mathbf{g}_2\right\|_2 & \cdots & \left\|\mathbf{g}_M\right\|_2\end{bmatrix}^T.
\label{ref:eq_udef}
\end{equation}
Using \eqref{ref:l0_equality}, the microphone subset selection problem for WPE can be defined as
\begin{equation}
\min_{\mathbf{g}}\left\| \mathbf{x_1} - \mathbf{X}_{\boldsymbol{\tau}}\mathbf{g} \right\|_p\quad \text { s.t. }\left\|\mathbf{u}\right\|_0 = K-1.
\label{ref:eq_l0norm}
\end{equation}
However, the optimization problem in \eqref{ref:eq_l0norm} is difficult to solve efficiently, both due to the nonconvexity of the $\ell_p$-norm for $0<p<1$ as well as the nonconvex $\ell_0$-norm constraint, which turns \eqref{ref:eq_l0norm} into a combinatorial problem.
One possible approach to reformulate the $\ell_p$-norm as a convex function is using the weighted $\ell_2$-norm \cite{Daubechies2010,Jukic2015}, leading to the following intermediate problem
\begin{equation}
\min_{\mathbf{g}}\left\| \mathbf{x_1} - \mathbf{X}_{\boldsymbol{\tau}}\mathbf{g} \right\|_{\mathbf{W}}^2\quad \text { s.t. }\left\|\mathbf{u}\right\|_0 = K-1
\label{ref:eq_l0wl2norm},
\end{equation}
where $\left\|.\right\|_{\mathbf{W}}$ denotes the weighted $\ell_2$-norm with the weighting matrix $\mathbf{W}$ typically updated iteratively for $I$ iterations. \par
A popular approach to solve combinatorial problems as in \eqref{ref:eq_l0wl2norm} efficiently is to perform convex relaxation \cite{Joshi2009}, whereby the $\ell_0$-norm constraint is replaced with a constraint on the $\ell_1$-norm. The motivation behind this step is that the $\ell_1$-norm has been shown to be the closest convex approximation to the $\ell_0$-norm \cite{Wipf2010}, therefore allowing the subset selection problem to be solved using conventional optimization methods. \par
We propose to reformulate the problem in \eqref{ref:eq_l0wl2norm} using convex relaxation, i.e.
\begin{equation}
\min_{\mathbf{g}}\left\| \mathbf{x_1} - \mathbf{X}_{\boldsymbol{\tau}}\mathbf{g} \right\|_{\mathbf{W}}^2\quad \text { s.t. }\left\|\mathbf{u}\right\|_1 = \sum_{m = 2}^M\left\|\mathbf{g}_m\right\|_2 = C,
\label{ref:eq_l1normeq}
\end{equation}
where $C$ is a constant. This problem can be alternatively formulated as \cite{Tibshirani1996} 
\begin{equation}
\boxed{\min_{\mathbf{g}}\underbrace{\left\| \mathbf{x_1} - \mathbf{X}_{\boldsymbol{\tau}}\mathbf{g} \right\|_{\mathbf{W}}^2}_{f(\mathbf{g})} + \underbrace{\lambda\sum_{m = 2}^M\left\|\mathbf{g}_m\right\|_2}_{h(\mathbf{g})}}
\label{ref:eq_groupsparsity}
\end{equation}
for an appropriate choice of the hyperparameter $\lambda$. The term $h(\mathbf{g})$  in \eqref{ref:eq_groupsparsity} is the well known group sparsity penalty \cite{Yuan2006}, also known as the $\ell_{2,1}$-norm. The group sparsity penalty $h(\mathbf{g})$ is nondifferentiable and convex, hence making the overall problem in \eqref{ref:eq_groupsparsity} a nondifferentiable convex problem. Typically, the group sparsity hyperparameter $\lambda$ is calculated as $\lambda = \lambda_{\text{c}}\lambda_{\text{max}}$ \cite{Parikh2014}, where the group sparsity factor $\lambda_{\text{c}}$ is a constant and $\lambda_{\text{max}}$ denotes a data-dependent maximum value. 
\subsection{Iterative optimization using proximal gradient}
Many different methods have been proposed to solve nondifferentiable convex optimization problems such as \eqref{ref:eq_groupsparsity}, one popular method being the proximal gradient algorithm \cite{Parikh2014}.
The proximal gradient algorithm, also known as the iterative shrinkage-thresholding algorithm (ISTA) and its accelerated version, the fast iterative shrinkage-thresholding algorithm (FISTA), are well suited for solving problems that can be decomposed into a convex differentiable and nondifferentiable part, i.e. $f(\mathbf{g})$ and $h(\mathbf{g})$ in \eqref{ref:eq_groupsparsity}, respectively. 
In each iteration, the proximal gradient algorithm combines a gradient descent step on $f(\mathbf{g})$ with the proximal operator of $h(\mathbf{g})$. The proximal operator can be viewed as a generalized projection and allows to efficiently minimize potentially nondifferentiable functions. \par
Applying the proximal gradient algorithm to the problem at hand in \eqref{ref:eq_groupsparsity} yields the following iterative solution for $\mathbf{g}$
\begin{equation}
\mathbf{g}^{(j + 1)} = \text{T}_m\left(\mathbf{g}^{(j)} - \alpha\left(\mathbf{X}_{\boldsymbol{\tau}}^H\mathbf{W}\mathbf{X}_{\boldsymbol{\tau}}\mathbf{g}^{(j)} - \mathbf{X}_{\boldsymbol{\tau}}^H\mathbf{W}\mathbf{x}_1\right),\lambda\right),
\label{eq:iteration_FISTA}
\end{equation}
where $j$ denotes the proximal gradient iteration index and $\alpha$ is the step-size. $\text{T}_m(\mathbf{g}_m)$ is an operator which includes the proximal mapping of the group sparsity penalty $h(\mathbf{g})$, given by
\begin{equation}
\text{T}_m(\mathbf{g}_m,\lambda) = 
\begin{cases}
  \mathbf{g}_m, & \text{if}\ m=1 \\
   \max\left(1 - \frac{\alpha\lambda}{\left\|\mathbf{g}_m\right\|_2},0\right)\mathbf{g}_m, & \text{otherwise}
\end{cases}.
\end{equation}
For the accelerated proximal gradient algorithm or FISTA\cite{Parikh2014}, an additional momentum term $\mathbf{y}$ is computed for each iteration in \eqref{eq:iteration_FISTA}, i.e.
\begin{equation}
\mathbf{y}^{(j)} = \mathbf{g}^{(j)} + \frac{j}{j + 3}\left(\mathbf{g}^{(j)} - \mathbf{g}^{(j - 1)}\right),
\end{equation}
replacing the update step in $\eqref{eq:iteration_FISTA}$ with
\begin{equation}
\mathbf{g}^{(j + 1)} = \text{T}_m\left(\mathbf{y}^{(j)} - \alpha\left(\mathbf{X}_{\boldsymbol{\tau}}^H\mathbf{W}\mathbf{X}_{\boldsymbol{\tau}}\mathbf{y}^{(j)} - \mathbf{X}_{\boldsymbol{\tau}}^H\mathbf{W}\mathbf{x}_1\right),\lambda\right).
\end{equation}
\subsection{Variable selection}
When performing convex relaxation, an additional step of variable selection is typically performed, e.g., in the form of a thresholding or maximum/minimum operation. To select a subset $\mathcal{S}$ of $K$ microphones out of the available $M$ microphones, we propose to select the microphones with the $K-1$ largest entries in the group vector $\mathbf{u}$ alongside the fixed reference microphone. Note that performing variable selection requires running the WPE algorithm once more on the selected subset. The complete proposed microphone subset selection method is outlined in Algorithm 1.\par
Since each frequency bin is processed independently, the selected microphone subsets $\mathcal{S}$ are inherently frequency-dependent. To select the same $K$ microphones for all frequency bins, frequency-independent microphone subset selection can be achieved by performing the variable selection step on the broadband group vector $\mathbf{u}_b = \sum_{f = 1}^F\mathbf{u}(f)$.
\begin{algorithm}
    \KwPar{Subset size $K$, sparsity-promoting parameter $p$,}
    group sparsity factor $\lambda_c$, number of reweighting iterations $I$, 

    number of accelerated proximal gradient iterations $J$,
    
    filter length $L_g$, microphone-dependent prediction delays $\tau_m$

    \KwIn{Microphone signals $\mathbf{x}_m$ $\forall m$ }

    \KwOut{Microphone subset $\mathcal{S}$}
    $\mathbf{x}_1 \leftarrow N\cdot\mathbf{x}_1 / \left\|\mathbf{x}_1\right\|_p$
    
    $\mathbf{x}_m(n - \tau_m) \leftarrow N\cdot\mathbf{x}_m(n - \tau_m)/\left\|\mathbf{x}_m(n - \tau_m)\right\|_p $
    
    $\mathbf{X}_{\tau_m} \leftarrow \text{compute convolution matrix using } \mathbf{x}_m(n - \tau_m)$
    $\mathbf{d}^{(1)} \leftarrow \mathbf{x}_1$
    
    \For{$i \leftarrow 1$ \KwTo $I$}{
        $\mathbf{W}^{(i)} \leftarrow \text{diag}(\lvert\mathbf{d}^{(i)}\rvert^{2} + 10^{-8})^{p/2 - 1}$ 
        
        $\mathbf{A} \leftarrow \mathbf{X}_{\boldsymbol{\tau}}^H\mathbf{W}^{(i)}\mathbf{X}_{\boldsymbol{\tau}}$
        
        $\mathbf{b} \leftarrow \mathbf{X}_{\boldsymbol{\tau}}^H\mathbf{W}^{(i)}\mathbf{x}_1$

        $\alpha \leftarrow 1/\text{P}(\mathbf{A})$
        
        $\lambda \leftarrow 2\lambda_c\left\|\mathbf{b}\right\|_{\infty}$

        \For{$j \leftarrow 1$ \KwTo $J$}{ 
            $\mathbf{y}^{(j)} \leftarrow \mathbf{g}^{(j)} + \frac{j}{j + 3}\left(\mathbf{g}^{(j)} - \mathbf{g}^{(j - 1)}\right)$
            
            $\mathbf{g}^{(j + 1)} \leftarrow \text{T}_m\left(\mathbf{y}^{(j)} - \alpha\left(\mathbf{A}\mathbf{y}^{(j)} - \mathbf{b}\right),\lambda\right)$
            
         }
         $\mathbf{d}^{(i + 1)} \leftarrow\mathbf{x}_1 - \mathbf{X}_{\boldsymbol{\tau}}\mathbf{g}^{(J)}$
    }
    $\mathcal{S} \leftarrow \{1,(K-1)\text{-max}(\mathbf{u})\}$
    \caption{Microphone subset selection using group sparsity penalty}
\end{algorithm}

\section{Experimental evaluation}
In this section, we evaluate the performance of the proposed frequency-dependent and frequency-independent microphone subset selection methods for an acoustic sensor network in a reverberant enclosure. In Section 4.1, we discuss the considered acoustic scenario and the algorithm parameters. In Section 4.2, we present the simulation results and evaluate the performance for different subset sizes.
\subsection{Acoustic setup and algorithm parameters}
\begin{figure}[t!]
\centering
\includegraphics[width=0.4\columnwidth]{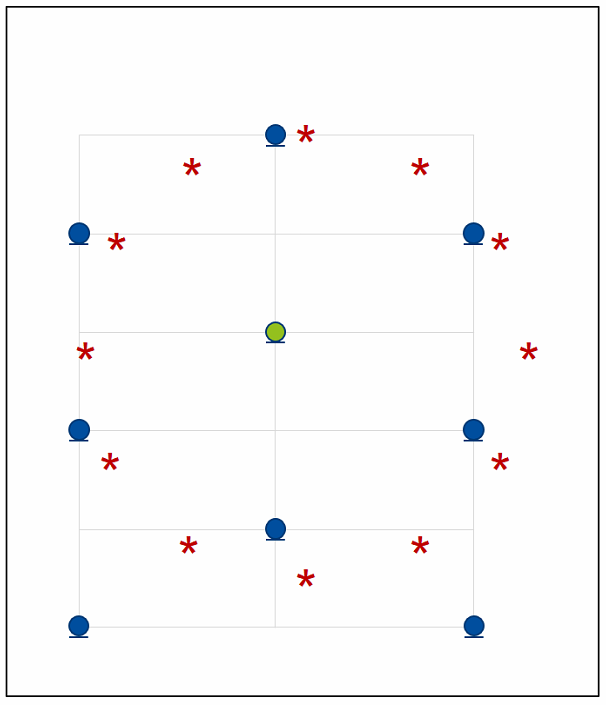}
\caption{Positions of $M=9$ spatially distributed microphones (\includegraphics[width=0.2cm]{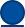}) with fixed reference microphone (\includegraphics[width=0.2cm]{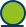}) and 12 considered speech source positions (\includegraphics[width=0.2cm]{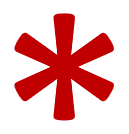})}
\label{fig:acoustic_scen}
\vspace{-0.55cm}
\end{figure}

We consider an acoustic sensor network with $M = 9$ spatially distributed microphones and a single speech source in a laboratory with dimensions of about $6$m$\times7$m$\times2.7$m and reverberation time $T_{60} \approx 1300$ ms. Fig. \ref{fig:acoustic_scen} depicts the position of the microphones and the considered positions of the speech source. The microphones are placed nonuniformly on a grid with of dimensions $4$m$\times5$m. The reference microphone is chosen as the microphone in the approximate center of the network and fixed for all considered source positions. In total 12 source positions are considered  on a circle with equal spacing between the source positions. \par
The reverberant microphone signals were generated at a sampling rate of $16$ kHz by convolving anechoic speech signals from the TIMIT database \cite{TIMIT} with measured room impulse responses from the BRUDEX database \cite{Fejgin2023}. The signals were processed using an STFT framework with frame size of $1024$ samples, frame shift $L_{\text{shift}} = 256$ samples and square-root Hann analysis and synthesis windows. The microphone-dependent prediction delays were estimated using the generalised cross-correlation with phase transform (GCC-PHAT) \cite{Knapp1976} and implemented using crossband filtering\cite{Lohmann2023}. \par 
The proposed microphone subset selection algorithm was implemented with the following WPE parameters: prediction filter length $L_g = 20$, prediction delay $\tau = 2$, sparsity-promoting parameter $p = 0.5$ and number of reweighting iterations $I = 10$. The group sparsity hyperparameter $\lambda$ was computed using a maximum value $\lambda_{\text{max}} = 2\left\|\mathbf{X}_{\boldsymbol{\tau}}^H\mathbf{W}\mathbf{x}_1\right\|_{\infty}$ \cite{Parikh2014} and we considered different group sparsity factors $\lambda_c \in \{10^{-5},10^{-4},10^{-3},10^{-2},10^{-1},1\}$. The accelerated proximal gradient algorithm was implemented with step-size $\alpha = 1/\text{P}(\mathbf{X}_{\boldsymbol{\tau}}^H\mathbf{W}\mathbf{X}_{\boldsymbol{\tau}})$ and number of iterations $J = 50$, where the operator $\text{P}(.)$ computes the largest eigenvalue.
\subsection{Simulation results}
First, in section 4.2.1, we select the value of the group sparsity factor $\lambda_c$ based on the WPE cost function in \eqref{ref:og_problem} when performing frequency-dependent microphone subset selection. Secondly, in section 4.2.2, using the selected group sparsity factor we evaluate the dereverberation performance of the processed signal using selected microphones. The dereverberation performance is measured using the perceptual evaluation of speech quality (PESQ). The reference signal used in PESQ is the direct component in the reference microphone.
\subsubsection{Frequency-dependent microphone subset selection}
For different values of the group sparsity factor $\lambda_c$, Fig. 2 depicts the difference between the average WPE cost function $J$ in \eqref{ref:og_problem} for the proposed frequency-dependent microphone subset selection method ($J^{\text{avg}}_{\text{GS}}$) and the optimal subset selection using exhaustive search ($J^{\text{avg}}_{\text{optimal}}$).
The average frequency-dependent cost functions have been computed by averaging over all frequency bins for all 12 considered source positions and different subset sizes $K \in \{2,3,4,5\}$. Hence the results in Fig. 2 can be seen as an overall measure of the performance of the proposed frequency-dependent subset selection method, where it can be seen that the best performance can be achieved when setting the group sparsity factor $\lambda_c = 10^{-2}$, as it minimizes both the mean cost difference and its standard error. \par
For the best and worst trial (combination of source position and subset size), Fig. 3 depicts the WPE cost per frequency for the optimal solution ($J_{\text{optimal}}$) and the proposed frequency-dependent microphone subset selection algorithm ($J_{\text{GS}}$) using a fixed group sparsity factor $\lambda_c = 10^{-2}$. For the best trial, it can be seen in Fig. 3a that the proposed method performs close to the optimal solution. For the worst trial, it can be seen in Fig. 3b that there is a larger difference between the performance of the proposed method and the optimal solution.
\subsubsection{Frequency-independent microphone subset selection}
Using the selected group sparsity factor, Fig. \ref{fig:fig4} depicts the average performance improvement over all considered source positions in terms of $\Delta$PESQ for the proposed frequency-independent microphone subset selection method. For different subset sizes $K \in \{2,3,4,5\}$, the performance of the proposed method using the group sparsity factor $\lambda_c = 10^{-2}$ is compared to the performance using the optimal exhaustive search solution based on \eqref{ref:eq_l0norm}, the performance using a randomly selected subset and the performance using all $M = 9$ microphones. First, it can be seen that the performance of the proposed method is close to that of the optimal solution for all considered subset sizes $K$. Secondly, using the proposed method the performance when using only 4 microphones is very close to that when using all 9 microphones.

\pagebreak
\begin{figure}[t!]
\centering
\includegraphics[width=0.6\columnwidth]{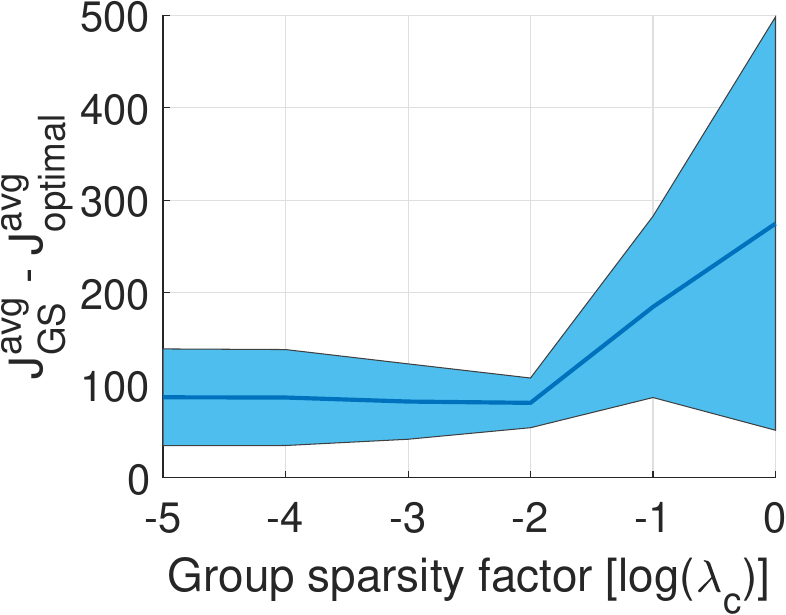}
\caption{Average WPE cost difference between proposed frequency-dependent method and the optimal solution for different values of the group sparsity factor $\lambda_c$}
\label{fig:fig2}
\vspace{-0.2cm}
\end{figure}%
\begin{figure}[t!]%
\centering
\begin{subfigure}[H]{0.48\columnwidth}
  \includegraphics[width=\columnwidth]{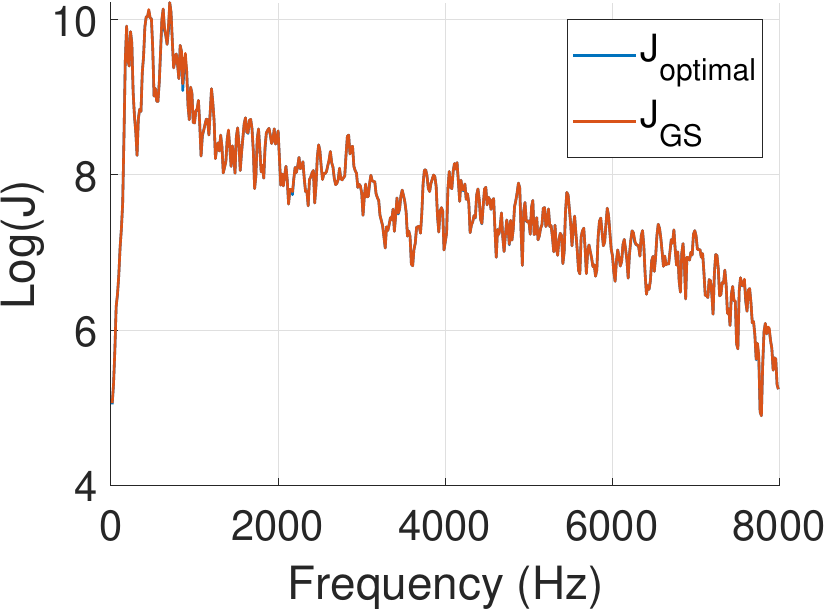}
  \vspace{-0.3cm}
  \caption{Best trial}%
  \label{subfig: a}%
\end{subfigure}\hfill%
\begin{subfigure}[H]{0.48\columnwidth}
  \includegraphics[width=\columnwidth]{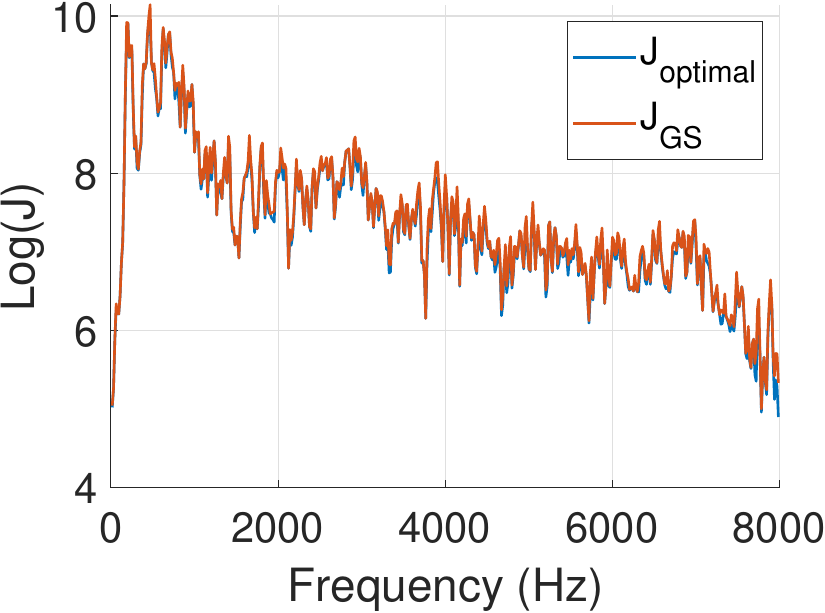}
  \vspace{-0.3cm}
  \caption{Worst trial}%
  \label{subfig: b}%
\end{subfigure}
\vspace{-0.2cm}
\caption{WPE cost for optimal solution and proposed frequency-dependent microphone subset selection using $\lambda_c = 10^{-2}$ for best and worst trials}
\label{fig:fig3}
\vspace{-0.2cm}
\end{figure}
\begin{figure}[t!]
\centering
\includegraphics[width=\columnwidth]{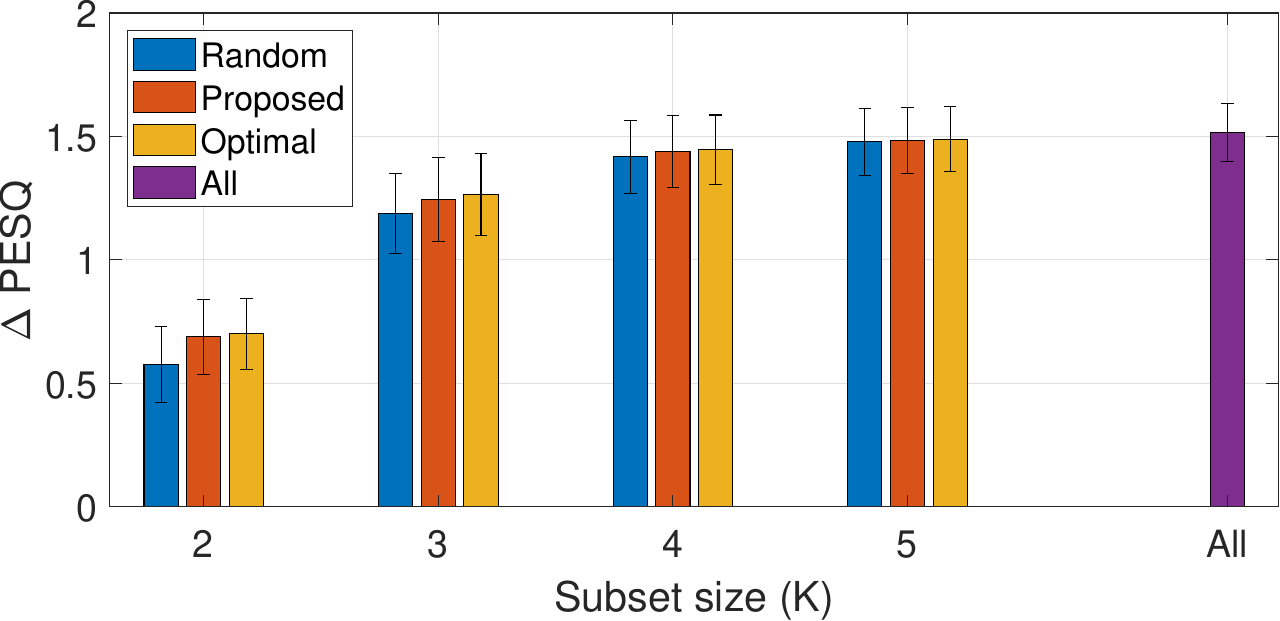}
\caption{Average PESQ improvement for proposed frequency-independent microphone subset selection algorithm, optimal solution and random selection for different subset sizes, compared to using all microphones}
\label{fig:fig4}
\vspace{-0.55cm}
\end{figure}%
\section{Conclusion}
In this paper we have presented a microphone subset selection method for the WPE algorithm. Using the popular convex relaxation method on the microphone subset selection problem, we performed microphone subset selection by introducing a group sparsity penalty on the prediction filter coefficients. Using measured impulse responses, we have evaluated the performance of the proposed frequency-dependent and frequency-independent microphone subset selection methods for a range of microphone subset sizes. The experimental evaluation showed that the performance of the proposed methods is close to the performance of the optimal exhaustive search approach using a fixed group sparsity factor. Furthermore, when using the proposed method, performance similar to using all 9 microphones can be achieved with only 4 microphones.




\bibliographystyle{IEEEbib}
\balance
\bibliography{refs}
\end{document}